\documentclass[pra, twocolumn, showpacs]{revtex4}
\usepackage{amsfonts}
\usepackage{amssymb}
\usepackage[dvips]{graphicx}
\usepackage{amsmath}
\usepackage{textcomp}

\newcommand{\Tr}{\textrm{Tr}}
\newcommand{\ket}[2][]{{|#2\rangle_{#1}}}

\newcommand{\OneThird}{\tfrac{1}{3}}

\newcommand{\neswarrow}{\mathrel{\text{$\nearrow$\llap{$\swarrow$}}}}
\newcommand{\nwsearrow}{\mathrel{\text{$\nwarrow$\llap{$\searrow$}}}}

\begin{document}

\title{Quantum and semiclassical polarization correlations}

\author{Konrad Banaszek}
\affiliation{Institute of Physics, Nicolaus Copernicus University, Grudziadzka 5, PL-87-100 Toru\'{n}, Poland}

\author{Rafa{\l} Demkowicz-Dobrza{\'n}ski}
\affiliation{Institute of Physics, Nicolaus Copernicus University, Grudziadzka 5, PL-87-100 Toru\'{n}, Poland}

\author{Micha{\l} Karpi{\'n}ski}
\affiliation{Institute of Experimental Physics, University of Warsaw, Ho\.{z}a 69, PL-00-681 Warsaw, Poland}

\author{Piotr Migda{\l}}
\affiliation{Institute of Experimental Physics, University of Warsaw, Ho\.{z}a 69, PL-00-681 Warsaw, Poland}

\author{Czes{\l}aw Radzewicz}
\affiliation{Institute of Experimental Physics, University of Warsaw, Ho\.{z}a 69, PL-00-681 Warsaw, Poland}

\begin{abstract}
We analyze the strength of polarization correlations between two light beams that can be achieved in the semiclassical regime using statistical mixtures of coherent states and binary on/off detectors. Under certain symmetry assumptions, the visibility of polarization correlations is shown to be bounded by $\OneThird$, which is in a striking contrast with perfect $100\%$ correlations exhibited by photon pairs prepared in the singlet state. The semiclassical limit is demonstrated in a measurement performed on a pair of laser beams undergoing correlated depolarization. This result illustrates the dramatic difference between predictions of quantum mechanics and the semiclassical theory of electromagnetic radiation for the polarization degree of freedom.
\end{abstract}

\maketitle

\section{Introduction}
Quantum mechanics allows for correlations that cannot be understood within the classical paradigm of local realism, which naturally grows out of experiences gathered in the macroscopic world \cite{WodkiewiczContPhys}. Models based on this paradigm are not able to reproduce fully predictions of the quantum theory unless supplied with additional controversial assumptions like non-locality, superluminal communication, or negative probabilities \cite{SuPRA91,WodkiewiczPRA95,WodkPRA95,BanaWodkPRA98}. A great deal of effort has been dedicated to identify the borders defined by local realistic theories, beyond which quantum mechanics offers the most dramatic manifestations of its unique character. This work currently finds applications in novel, quantum-based protocols for secure communication and information processing \cite{EkertPRL91,DenseCoding,AcinQKD}.

One of the major sources of excitement in quantum optics has been tracing effects that can be explained only on the grounds of the fully quantum theory of light and matter. A standard reference point here is the semiclassical theory of photodetection, in which a quantized material system interacts with light treated classically. This theory suffices to explain a wealth of optical observations in systems comprising stochastic classical fields, linear optical elements, and typical photodetectors \cite{MandelJOSA77}. A direct demonstration of the quantum nature of light requires one to resort to more exotic ways to generate and manipulate light \cite{KimblePRL77}. The semiclassical regime can be viewed as a special case of a local realistic theory, and its bounds are typically tighter, as discussed in detail for two-photon interference effects \cite{SuPRA91}. However, the semiclassical regime can be easily implemented in an optical laboratory using readily available resources, unlike other models considered in general local realistic theories which may include highly speculative assumptions.

When identifying the limits of the semiclassical regime, it is crucial to take into account a realistic description of the employed components. For example, most detectors with single photon sensitivity do not have photon number resolution. As a result, intensity correlations typical for non-classical sources of optical radiation can be mimicked by suitably engineered non-stationary classical fields, and additional tests of quantumness are needed \cite{URenPRA05}. A thorough study of the actual construction details of standard photon counting modules can even provide a complete control over their behaviour using classical light \cite{MakarovNJP09}, which leads to new eavesdropping strategies for quantum cryptography \cite{MakarovXXX}. This clearly shows that understanding the precise borders between semiclassical, local realistic, and quantum regimes remains full of challenges that grow out of previous advances in quantum optics and are highly relevant to emerging quantum-enhanced technologies.

The purpose of this paper is to explore the question to what degree the semiclassical regime can reproduce polarization correlations. Such correlations are a celebrated signature of quantumness \cite{Bohm} and nowadays can be routinely observed for photon pairs produced in a suitably arranged process of parametric down-conversion \cite{KwiatPRL95}. Typical for most quantum optical experiments, we will assume here that the photodetectors provide only a binary response telling whether at least one photon has been detected or none at all \cite{Mogilevtsev,NoPhotonRes,BanaWodkPRL99,BanaDragPRA02}. The probability of a click is a non-linear function of the incident intensity. This can be viewed as a generalization of the Malus law \cite{WodkiewiczPRA95} considered within the semiclassical theory. The theoretical discussion is illustrated with a photon counting measurement performed on a pair of collectively depolarized light beams that provide the strongest form of polarization correlations allowed in the semiclassical regime.

The quest to understand the exact relation between the classical and the quantum has been vigorously pursued by and will remain an important element of the scientific legacy of late Krzysztof W\'{o}dkiewicz. This contribution is dedicated to his memory.

\section{Quantum correlations}

The notion of correlations in classical physics is reserved to the statistical description of systems for which complete knowledge is \emph{not available}. The state of a system is said to be correlated whenever the joint probability distribution describing parameters of the entire system does not factorize into marginal probability distributions for individual subsystems. If the complete knowledge about the system is available, the probability distribution is clearly a product
of marginal probability distributions, since the state of each subsystems is definite and hence no correlations are present.

According to the standard interpretation of quantum mechanics, the complete knowledge of a quantum system is encoded
in a vector $\ket{\psi}$ in an appropriate Hilbert space. The superposition principle together with the tensor product
structure of composite states implies the existence of \emph{entangled states}---a unique quantum concept
comprising the complete knowledge about the composite system and correlations within.
The strength of correlations present in entangled states may well outperform any conceivable classical ones and
therefore they provide an argument for the incompatibility of local realistic theories with the quantum world via Bell's inequalities \cite{Bell1964}.

One of the simplest and at the same time the most characteristic examples of an entangled state is the \emph{singlet state}. When realized in the polarization degree of freedom of two photons, the singlet state reads:
\begin{equation}
\label{eq:singlet}
\ket{\Psi^-}=\frac{1}{\sqrt{2}}\bigl(\ket{\mathord{\leftrightarrow}}_a\otimes
\ket{\mathord{\updownarrow}}_b - \ket{\mathord{\updownarrow}}_a\otimes \ket{\mathord{\leftrightarrow}}_b\bigr),
\end{equation}
where the kets $\ket{\mathord{\leftrightarrow}}$ and $\ket{\mathord{\updownarrow}}$ correspond to the horizontal and the vertical polarization states respectively, and the indices $a$ and $b$ refer to the two photons that are assumed to be distinguishable by another degree of freedom, for example the direction of propagation.
A feature that distinguishes the singlet state from other two-photon entangled states is its invariance under
arbitrary polarization transformations, provided that the same transformation is applied to both the photons.
Consequently, the anticorrelations present in the horizontal-vertical basis in Eq.~(\ref{eq:singlet})
manifest themselves also in any other basis the polarization is being measured.

More generally, let ${\bf u}_a$ and  ${\bf u}_b$ be two Bloch vectors of unit length representing the polarization basis in which the polarizations are being measured on the two photons respectively. It is easy to verify that the density operator corresponding to the singlet state can be written as a special case of a two-qubit Werner state \cite{WernerState}
\begin{equation}
\label{Eq:varrho=sigma}
\hat{\varrho} = \frac{1}{4} \left( \hat{\openone} \otimes \hat{\openone} +
\eta \sum_{i=x,y,z} \hat{\sigma}_i \otimes \hat{\sigma}_i \right),
\end{equation}
with $\eta = -1$. In the above formula, $\hat{\sigma}_i$ are Pauli operators. This representation allows us to calculate in a straightforward way, using trace properties of Pauli operators, the joint probability of measuring the two photons in respective polarizations ${\bf u}_a$ and ${\bf u}_b$:
\begin{multline}
p({\bf u}_a, {\bf u}_b)= \frac{1}{4}\Tr[ \hat{\varrho}
(\hat{\openone} + {\bf u}_a \cdot \hat{\boldsymbol{\sigma}})\otimes
(\hat{\openone} + {\bf u}_b \cdot \hat{\boldsymbol{\sigma}} )] \\
= \frac{1}{4}[1 + \eta {\bf u}_a \cdot {\bf u}_b ]
\end{multline}
where $\hat{\boldsymbol{\sigma}}$ denotes a three-component vector composed of the Pauli operators.
Thus the joint probability for polarization settings ${\bf u}_a$ and  ${\bf u}_b$ is a function of the angle $\alpha$ between the corresponding Bloch vectors, given by $\cos\alpha = {\bf u}_a \cdot {\bf u}_b$. When performing a \emph{common}
transformation on both measured polarizations ${\bf u}_a$ and ${\bf u}_b$, the probability is left unchanged.
It is seen that the visibility of polarization correlations is given by $|\eta|$, reaching $100\%$ for the singlet state. The density operator defined in Eq.~(\ref{Eq:varrho=sigma}) is positive definite for $-1 \le \eta \le \tfrac{1}{3}$. Thus although perfect {\em anticorrelations} in the polarization degree of freedom are possible, there is no physical quantum state in which the two photons would always have identical polarizations in an arbitrary basis.

\section{Semiclassical bounds}

Let us consider a pair of light beams distributed between two parties $a$ and $b$. Because of the employed detection scheme, it will be sufficient to take into account only the intensity and the polarization degrees of freedom.
Classically, a general state of polarization is described by a four-component Stokes vector. For a fully polarized beam the component corresponding to the total intensity is unambiguously defined by the remaining three coordinates. In order to retain correspondence with the quantum mechanical case, we will order these coordinates identically as in the Bloch vectors and we will call here the resulting three-component objects {\em generalized} Bloch vectors.

Thus an individual preparation of the two beams is described by generalized Bloch vectors
${\bf s}_a$ and ${\bf s}_b$, whose length corresponds to the intensity, and the direction---to the polarization state of the respective beam. If the complete knowledge about the beams is unavailable,
the joint state is described by a probability distribution $P({\bf s}_a,{\bf s}_b)$.
In the following, we shall assume that the state is invariant with respect to an arbitrary common polarization transformation $P({\boldsymbol\Omega}{\bf s}_a, {\boldsymbol\Omega}{\bf s}_b) = P({\bf s}_a, {\bf s}_b)$, where $\boldsymbol{\Omega}$ belongs to the SO(3) group of proper rotations in three real dimensions.

\begin{figure}[t]
\includegraphics[width=1 \columnwidth]{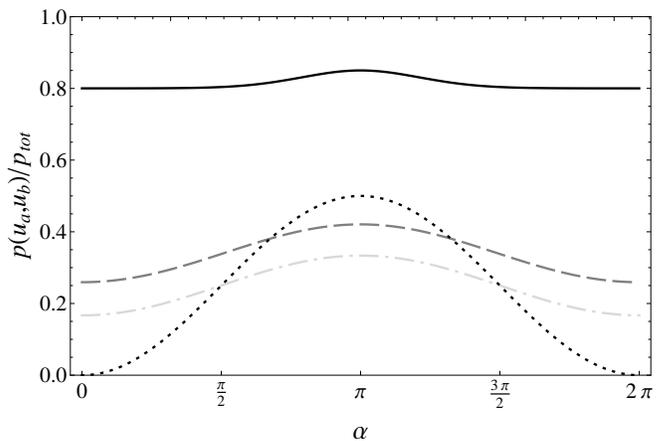}
\caption{The coincidence probability as a function of the angle $\alpha$ between the Bloch vectors of measured polarizations for the singlet state (dotted, black) and a rotationally invariant classical state of two beams with perfectly anticorrelated polarizations and identical intensities, equal to $s=0.0025$ (dashed-dotted, light gray), $s=1$ (dashed, dark gray), and $s=10$ (solid, black). The curves are normalized by
the probability $p_{\text{tot}}$ of a coincidence event when entire beams without selecting polarization components
are detected.}
\label{fig:coinc}
\end{figure}

The measurement is performed by selecting polarization components corresponding to ${\bf u}_a$ and ${\bf u}_b$ in the respective beams, and detecting clicks using single photon counting modules. Assuming that
the detector fires whenever one or more photons are detected, the click probability for the classical light of normalized intensity $s$ entering the detector reads
$\Pi(s) = 1 - \exp(-s)$ \cite{Photodetection}.
Therefore, the probability of a coincidence event in the semiclassical model is given by
\begin{multline}
p({\bf u}_a, {\bf u}_b) = \\
\int d{\bf s}_a \int d{\bf s}_b \, P({\bf s}_a, {\bf s}_b)
\Pi\bigl({\textstyle\frac{1}{2}} (|{\bf s}_a| + {\bf u}_a \cdot {\bf s}_a)\bigr)
\Pi\bigl({\textstyle\frac{1}{2}} (|{\bf s}_b| + {\bf u}_b \cdot {\bf s}_b)\bigr)
\end{multline}
Detector losses are described by multiplying the argument of $\Pi(s)$ by the detection efficiency, which can be included in the above formula by appropriate rescaling of the probability $P({\bf s}_a, {\bf s}_b)$.
Let us now represent ${\bf s}_a = s_a {\boldsymbol\Omega}{\bf e}$, where ${\bf e}$ is a reference unit vector, ${\boldsymbol\Omega}$ is a certain rotation, and $s_a = |{\bf s}_a|$.
Writing the integration measure
as $d{\bf s}_a = ds_a d\boldsymbol\Omega$ and making use of the assumed
rotational invariance of $P({\bf s}_a, {\bf s}_b)$
allows us to write
\begin{multline}
p({\bf u}_a, {\bf u}_b) = \int ds_a \int d{\boldsymbol\Omega} \int d{\bf s}_b \,
P( s_a {\bf e}, {\boldsymbol\Omega}^{-1}{\bf s}_b) \\ \times
\Pi\bigl({\textstyle\frac{1}{2}}s_a (1 + {\bf u}_a \cdot {\boldsymbol\Omega}{\bf e} ) \bigr)
\Pi\bigl({\textstyle\frac{1}{2}}(|{\bf s}_b| + {\bf u}_b \cdot {\bf s}_b ) \bigr).
\end{multline}
Substituting ${\bf s}_b \rightarrow {\boldsymbol\Omega} {\bf s}_b$ yields:
\begin{multline}
\label{eq:probsc}
p({\bf u}_a, {\bf u}_b) = \int ds_a \int d{\bf s}_b \,
P(s_a {\bf e}, {\bf s}_b)\\
\int d{\boldsymbol\Omega} \,
\Pi\bigl({\textstyle\frac{1}{2}}s_a (1 + {\bf u}_a \cdot {\boldsymbol\Omega}{\bf e} ) \bigr)
\Pi\bigl({\textstyle\frac{1}{2}}(|{\bf s}_b| + {\bf u}_b \cdot {\boldsymbol\Omega} {\bf s}_b ) \bigr).
\end{multline}
It will be convenient to assume in the following that the measure over ${\boldsymbol\Omega}$ is normalized to one,
$\int d{\boldsymbol\Omega} = 1$.
Using the explicit form of the count probability $\Pi(s)$, the integral over ${\boldsymbol\Omega}$
in Eq.~(\ref{eq:probsc}) can be simplified
to
\begin{multline}
\label{eq:probcu}
\int d{\boldsymbol\Omega} \,
\Pi\bigl({\textstyle\frac{1}{2}}s_a(1 + {\bf u}_a \cdot {\boldsymbol\Omega}{\bf e} ) \bigr)
\Pi\bigl({\textstyle\frac{1}{2}}(|{\bf s}_b| + {\bf u}_b \cdot {\boldsymbol\Omega} {\bf s}_b ) \bigr)
= \\
1 - \frac{1}{|{\bf s}_a|}(1-e^{-|{\bf s}_a|})- \frac{1}{|{\bf s}_b|}(1-e^{-|{\bf s}_b|})
+ e^{-(|{\bf s}_a|+|{\bf s}_b|)/2}\,  \mathcal{U},
\end{multline}
where the only term depending on the polarizations ${\bf u}_a$ and ${\bf u}_b$ being measured reads
\begin{equation}
\label{Eq:Udef}
\mathcal{U} =  \int d{\boldsymbol\Omega} \, e^{-(s_a {\bf u}_a \cdot {\boldsymbol\Omega}{\bf e}
+ {\bf u}_b \cdot {\boldsymbol\Omega}{\bf s}_b)/2}.
\end{equation}
This integral can be calculated explicitly when ${\bf s}_a = -{\bf s}_b$, leading to a closed expression for
the coincidence probability $p({\bf u}_a, {\bf u}_b)$ of the form
\begin{equation}
\label{eq:coinccurve}
p({\bf u}_a, {\bf u}_b) =   1 + \frac{2}{s} \left[
 e^{-s}
\left( 1 + \frac{\sinh(s\sin\frac{\alpha}{2})}{2\sin\frac{\alpha}{2}} \right) - 1 \right],
\end{equation}
where $\alpha$ is the angle between the vectors ${\bf u}_a$ and ${\bf u}_b$ and
$s= |{\bf s}_a| = |{\bf s}_b|$. In Fig.~\ref{fig:coinc} we depict
the dependence of the $p({\bf u}_a, {\bf u}_b)$ on the angle $\alpha$ for several values of the intensity $s$. In order to aid comparison, the graphs have been rescaled by the the probability $p_{\text{tot}}$ of a coincidence event when the polarizers are removed from the setup. It is seen than the correlations are much weaker than in the quantum mechanical case of a singlet state, and for larger intensities they loose the purely harmonic form. Note that choosing ${\bf s}_a = {\bf s}_b$ in Eq.~(\ref{Eq:Udef}) leads to an expression analogous to Eq.~(\ref{eq:coinccurve}) with $\sin\frac{\alpha}{2}$ replaced by $\cos\frac{\alpha}{2}$.

\begin{figure}[t]
\includegraphics[width=1 \columnwidth]{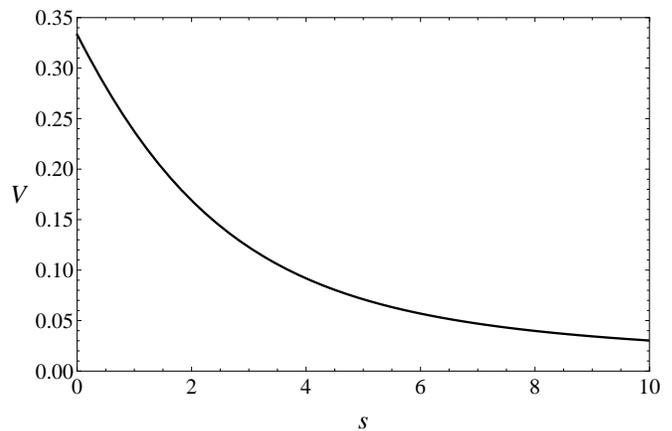}
\caption{The visibility of polarization correlations in the semiclassical model when both beams have equal intensities $|{\bf s}_a|=|{\bf s}_b|=s$ and either identical or orthogonal polarizations.}
\label{fig:visibility}
\end{figure}

\begin{figure}[t]
\includegraphics[width=0.9 \columnwidth]{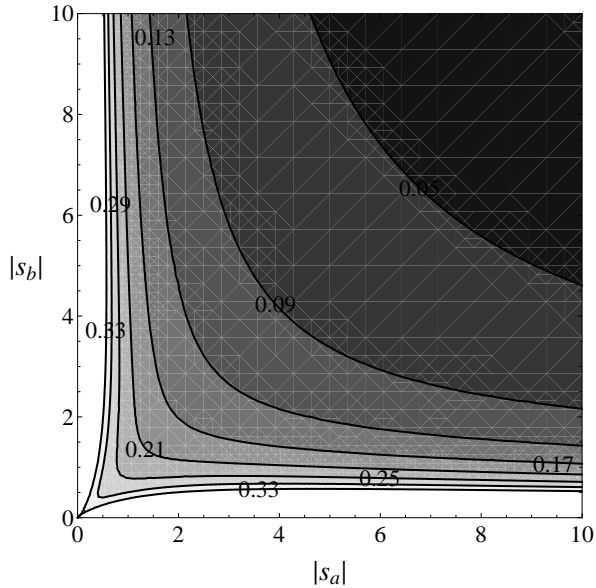}
\caption{A bound on the visibility of polarization correlations achievable using classical light, depicted as a function of the beam intensities $|{\bf s}_a|$ and  $|{\bf s}_b|$. The bound is tight for $|{\bf s}_a|=|{\bf s}_b|$. In the white regions near axes the bound exceeds the value $\OneThird$, conjectured to be the general visibility limit in the semiclassical regime.}
  \label{fig:boundplot}
\end{figure}

We will now derive bounds on $\mathcal{U}$ for arbitrary generalized Bloch vectors ${\bf s}_a$ and ${\bf s}_b$, which will be used to analyze the visibility of polarization correlations in a more general case. In order to find a lower bound
on $\mathcal{U}$ let us start from an observation that the integral in Eq.~(\ref{Eq:Udef}) is
invariant under the substitution
$\boldsymbol{\Omega} \rightarrow \boldsymbol{\Omega} \boldsymbol{\Omega}_0$. Taking $\boldsymbol{\Omega}_0$ to be
a $\pi$ rotation around the axis perpendicular to the plane spanned by ${\bf e}$ and ${\bf u}_b$ implies
that the value of $\mathcal{U}$ does not change under simultaneous substitutions ${\bf e} \rightarrow -{\bf e}$ and ${\bf u}_b \rightarrow - {\bf u}_b$. This allows us to write:
\begin{multline}
\mathcal{U} = \frac{1}{2} \int d \boldsymbol{\Omega}\bigl{(} e^{-( s_a {\bf u}_a \cdot {\boldsymbol\Omega}{\bf e}
+ {\bf u}_b \cdot {\boldsymbol\Omega}{\bf s}_b)/2} +  e^{( s_a {\bf u}_a \cdot {\boldsymbol\Omega}{\bf e}
+ {\bf u}_b \cdot {\boldsymbol\Omega}{\bf s}_b)/2}   \bigr{)} \\
= \int d \boldsymbol{\Omega} \cosh (| s_a {\bf u}_a \cdot {\boldsymbol\Omega}{\bf e}
+ {\bf u}_b \cdot {\boldsymbol\Omega}{\bf s}_b|/2).
\end{multline}
Using the convexity of the hyperbolic cosine function and the inverse triangle inequality we obtain
\begin{multline}
\mathcal{U} \ge  \cosh \left( \int d \boldsymbol{\Omega}| s_a {\bf u}_a \cdot {\boldsymbol\Omega}{\bf e}
+ {\bf u}_b \cdot {\boldsymbol\Omega}{\bf s}_b|/2 \right) \\
\ge \cosh \left( \int d \boldsymbol{\Omega}( | s_a {\bf u}_a \cdot {\boldsymbol\Omega}{\bf e} |
- | {\bf u}_b \cdot {\boldsymbol\Omega}{\bf s}_b| )/2\right).
\end{multline}
Evaluating the two integrals over $\boldsymbol{\Omega}$ yields
\begin{equation}
\label{Eq:Umin}
\mathcal{U} \ge \cosh[(|{\bf s}_a | - |{\bf s}_b|)/4].
\end{equation}
An upper bound on $\mathcal{U}$ can be found with the help of the Cauchy-Schwarz inequality
\begin{multline}
\mathcal{U}
 \leq
\sqrt{\left( \int d \boldsymbol{\Omega} e^{- s_a {\bf u}_a \cdot {\boldsymbol\Omega}{\bf e}} \right)
\left( \int d \boldsymbol{\Omega} e^{ -{\bf u}_b \cdot {\boldsymbol\Omega}{\bf s}_b}  \right) } \\
= \sqrt{\frac{\sinh |{\bf s}_a| \sinh |{\bf s}_b|}{|{\bf s}_a||{\bf s}_b|}}.
\label{Eq:Umax}
\end{multline}
The two derived bounds  allow us to estimate from above and from below the coincidence probability
$p({\bf u}_a, {\bf u}_b)$ over all possible settings of ${\bf u}_a$ and ${\bf u}_b$.

\section{Correlations visibility}

Let us first discuss the limit of weak intensities $|{\bf s}_b|, |{\bf s}_b| \ll 1$, when the probability of a click on a detector can be approximated by $\Pi(s) \approx s$. In this case, the probability of a coincidence event can be written as
\begin{equation}
p({\bf u}_a, {\bf u}_b) \approx \frac{1}{4} \left(
\bigl\langle |{\bf s}_a| |{\bf s}_b| \bigr\rangle + \frac{1}{3} \bigl\langle {\bf s}_a \cdot {\bf s}_b \bigr\rangle
\cos\alpha
\right)
\end{equation}
where the angular brackets denote the average $\langle \ldots \rangle = \int d{\bf s}_a  \int d{\bf s}_b
P({\bf s}_a, {\bf s}_b) \ldots$. The Cauchy-Schwarz inequality means that $\bigl| \bigl\langle {\bf s}_a \cdot {\bf s}_b \bigr\rangle \bigr| \le \bigl\langle |{\bf s}_a| |{\bf s}_b| \bigr\rangle$, which implies that the visibility cannot exceed $\OneThird$.

In a symmetric scenario, when both the beams always have orthogonal polarizations and equal intensities $|{\bf s}_a|=|{\bf s}_b|=s$ ,
we can use Eq.~(\ref{eq:coinccurve}) to calculate the visibility of polarization correlations
\begin{equation}
V = \frac{\sinh s - s}{s+(2 s-3) \sinh s+2 (s-2) \cosh s+4}.
\end{equation}
This quantity, shown in Fig.~\ref{fig:visibility}, is a monotonically decreasing function of the intensity $s$, with the maximum value $V=\OneThird$ reached in the limit $s\rightarrow 0$. Note that in this scenario the bounds given in Eqs.~(\ref{Eq:Umin}) and (\ref{Eq:Umax}) are tight.

\begin{figure}[t]
\includegraphics[width=0.8 \columnwidth]{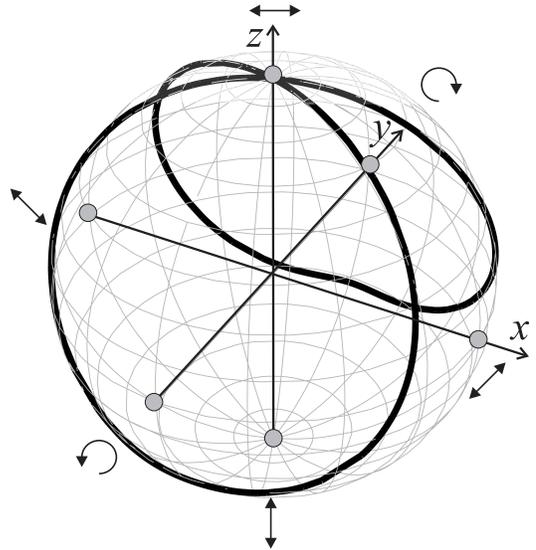}
\caption{The path drawn by the Bloch vector of a linearly polarized beam sent through a pair of counter-rotating waveplates with phase delays $\pi$ and $\arccos\frac{1}{\sqrt{3}}$. This transformation applied to two beams with initially orthogonal polarizations produces an ensemble equivalent in the limit of very weak intensities to a rotationally invariant one.}
  \label{fig:travel}
\end{figure}

More generally, we can use the inequalities derived in Eqs.~(\ref{Eq:Umin}) and (\ref{Eq:Umax}) to find an upper bound on the visibility of polarization correlations for arbitrary intensities $|{\bf s}_a|$ and $|{\bf s}_b|$. This bound is depicted in Fig.~\ref{fig:boundplot}.
It is seen that the visibility indeed does not exceed $\OneThird$ except regions when one of the intensities  $|{\bf s}_a|$ or $|{\bf s}_b|$ is small. Let us have a closer look at the limit when one of the intensities approaches zero. For concreteness, let us take $|{\bf s}_a| \rightarrow 0$. It is then justified to expand
\begin{equation}
\mathcal{U} \approx  \int d{\boldsymbol\Omega} \, (1- \tfrac{1}{2} {\bf u}_a \cdot {\boldsymbol\Omega}{\bf s}_a )
e^{-{\bf u}_b \cdot {\boldsymbol\Omega}{\bf s}_b/2}.
\end{equation}
This expression can be evaluated analytically to yield
\begin{multline}
\mathcal{U} \approx \frac{2}{|{\bf s}_b|} \sinh \frac{|{\bf s}_b|}{2}
+ \frac{1}{|{\bf s}_b|^2} ( {\bf u}_a \cdot {\bf u}_b ) ( s_a {\bf e} \cdot {\bf s}_b )
\\
\times \left( \cosh  \frac{|{\bf s}_b|}{2} - \frac{2}{|{\bf s}_b|} \sinh \frac{|{\bf s}_b|}{2} \right).
\end{multline}
\begin{figure}[t]
\includegraphics{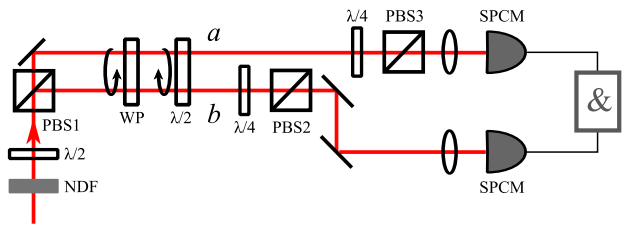}
\caption{The experimental setup. NDF, a set of neutral density filters; $\lambda/2$, $\lambda/4$, half and quarter wave plates; WP, a wave plate introducing an  $\arccos\tfrac{1}{\sqrt{3}}$~rad phase shift; PBS, polarizing beam splitters; SPCM, single photon counting modules.}
  \label{fig:setup}
\end{figure}

\begin{figure}[t]
\includegraphics[width=0.8 \columnwidth]{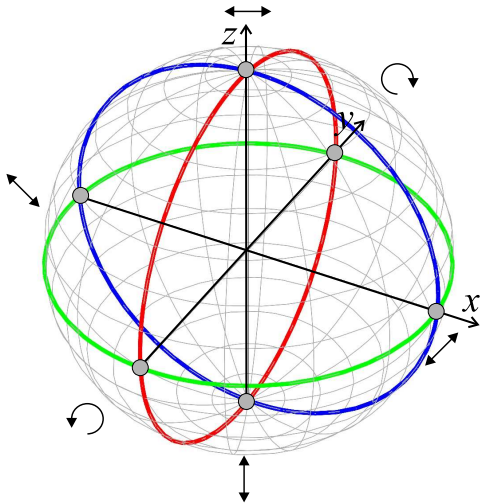}
\caption{The three great circles on the Bloch sphere specifying the polarization scanned for the beam $b$ to verify polarization correlations. The blue circle in the $xz$ plane was scanned for the horizontal $\leftrightarrow$ and the vertical $\updownarrow$ polarizations of the beam $a$. The red circle in the $yz$ plane was scanned for circular polarizations $\circlearrowright$ and $\circlearrowleft$ of the beam $a$. The green circle in the $xy$ plane was scanned for diagonal polarizations $\neswarrow$ and $\nwsearrow$ of the beam $a$.}
  \label{fig:circles}
\end{figure}
It is seen that the minimum and the maximum values of the coincidence probability are obtained when the product
$( {\bf u}_a \cdot {\bf u}_b ) ( s_a {\bf e} \cdot {\bf s}_b ) = \pm |{\bf s}_a| |{\bf s}_b|$. It can be verified by a straightforward yet lengthy calculation that the visibility of polarization correlations obtained using this expression approaches $\OneThird$ when $|{\bf s}_a| \rightarrow 0$. This shows that the upper bound derived with the help of the inequalities given in Eqs.~(\ref{Eq:Umin}) and (\ref{Eq:Umax}) is not tight and suggests that the limit $\OneThird$ might be universal for arbitrary intensities $|{\bf s}_a|$ and $|{\bf s}_b|$.

Let us note that the bound $\OneThird$ does not discriminate between positive and negative polarization correlations. The quantum mechanical counterpart a statistical ensemble discussed above is a situation when each of the two photons has a well defined polarization state, and only statistical averaging is permitted, i.e.\ the pair is prepared in a {\em separable state}. For Werner states introduced in Eq.~(\ref{Eq:varrho=sigma}) the separability criterion takes a form symmetric with respect to the sign of the parameter $\eta$: $-\tfrac{1}{3} \le \eta \le \tfrac{1}{3}$  \cite{HorodeckiPRA96}, also leading to the visibility limit of $\OneThird$ for polarization correlations.

\section{Experiment}

\begin{figure}[t]
\centering
\includegraphics{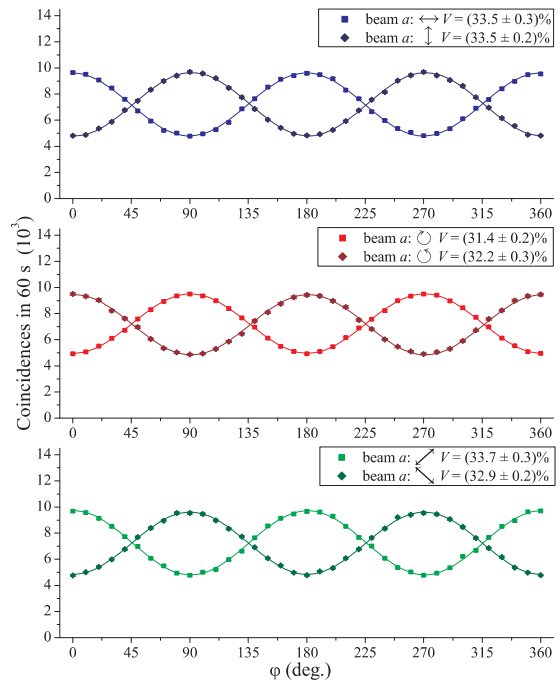}
\caption{The coincidence count rates as a function of the rotation angle $\varphi$ of the polarizer PBS2 placed in the path of the beam $b$. The three panels display results for scans of the polarization of the beam $b$ along the great circles in the $xz$ plane (top), the $yz$ plane (middle), and the $xy$ plane (bottom). The solid lines depict least-squares sinusoidal fits. The insets specify the polarization of the beam $a$ for each of the datasets along with visibilities obtained from the fits.}  \label{fig:results}
\end{figure}

The semiclassical limit of polarization correlations can be demonstrated experimentally using a pair of orthogonally polarized light beams, submitted to identical polarization transformations selected randomly according to the Haar measure on the rotation group SO(3). Mechanical implementation \cite{KarpRadzJOSAB08} of such depolarization would lead to excessively long averaging times. However, for low intensities, when the probability of a count is proportional to the incident intensity, the polarization measurement is sensitive only to certain moments of the distribution of output polarizations:
\begin{multline}
p({\bf u}_a, {\bf u}_b) \approx \tfrac{1}{4} \left( \bigl\langle |{\bf s}_a| |{\bf s}_a| \bigr\rangle
+ {\bf u}_a \cdot \bigl\langle |{\bf s}_a| {\bf s}_b \bigr\rangle
+ {\bf u}_b \cdot \bigl\langle {\bf s}_a |{\bf s}_b| \bigr\rangle \right. \\
\left. + {\bf u}_a \cdot \bigl\langle {\bf s}_a^T {\bf s}_b \bigr\rangle {\bf u}_b
\right)
\end{multline}
These moments can be reproduced by a discrete set of Bloch vectors, or a continuous one-parameter family. The latter solution is a convenient choice for an implementation in an optical setup. An exemplary realization is a set of two waveplates with phase delays equal respectively to $\arccos\tfrac{1}{\sqrt{3}}$~rad and $\pi$~rad that rotate in opposite directions with the same constant angular velocities.
For a suitably selected input linear polarization the output Bloch vector draws a curve depicted in Fig.~\ref{fig:travel}, which yields the output ensemble with desired statistical properties.

The complete experimental setup is presented in Fig.~\ref{fig:setup}. The light source was a highly attenuated pulsed beam from a mode-locked Ti:sapphire oscillator with the central wavelength  $775~\mathrm{nm}$ and the full width at half-maximum bandwidth $10~\mathrm{nm}$, corresponding to subpicosecond pulse duration. The polarizing beam splitter PBS1 followed by a silver mirror was used to prepare two parallel beams in horizontal $\leftrightarrow$ and vertical $\updownarrow$ polarizations with respect to the plane of the setup. The half wave plate inserted before PBS1 served to equalize the intensities of the two beams. The two wave plates introducing depolarization were mounted on rotation stages driven by stepper motors.  In order to produce the correct output ensemble, the axes of the waveplates were initially aligned parallel to each other. For each measurement point, the waveplates were rotated in opposite directions with constant angular velocities, completing an integer number of 10 full rotations.
The two emerging beams were transmitted through separate quarter wave plates $\lambda/4$ followed by polarizers, which selected individual polarization components, and then focused on the active areas of single photon counting modules SPCM (model Perkin Elmer SPCM-AQ-131). The input beam was attenuated to an intensity at which single count rates on the SPCMs with removed polarizers were approximately equal to $2 \times 10^5$~Hz. Given the laser repetition rate of $80~\mathrm{MHz}$ this yields the intensity parameter $s \approx 2.5 \times 10^{-3}$.

In order to verify the correlations, we chose six different polarizations for the beam $a$: horizontal $\leftrightarrow$, vertical $\updownarrow$, diagonal at $\pm 45$ degrees $\neswarrow$ and $\nwsearrow$, as well as right- and left-circular $\circlearrowright$ and $\circlearrowleft$. For each of these six polarizations, we measured coincidence count rates as a function of the polarization selected for the beam $b$ sweeping one of three great circles on the Bloch sphere, shown in Fig.~\ref{fig:circles}. The great circles were scanned by setting the quarter wave plate in the beam $b$ to an the appropriate orientation or removing it altogether for the scan of the $xz$ circle, and rotating the polarizer PBS2 in 10\textdegree\ steps, which corresponds to 20\textdegree\ increments on a great circle. For each setting, photocounts were collected over a $60~\mathrm{s}$ time interval, and both single and coincidence count numbers were recorded.

The measured polarization correlations are depicted in Fig.~\ref{fig:results}. The data sets were fitted using the least-squares method with sinusoidal patterns to obtain corresponding visibilities, specified in the graph insets.
The experimentally determined values are close to the predicted value of $\OneThird$. Small deviations can be attributed to non-uniform depolarization of the two beams. This imperfection could also be noticed as weak dependence of single count rates on the measured polarization, with relative variations up to $1.5\%$.

\section{Conclusions}

We investigated the strength of polarization correlations that can be achieved using classical optical fields and detectors without photon number resolution. We showed under certain symmetry assumptions that there is a substantial gap between the semiclassical regime and the correlations allowed by quantum mechanics. It is worthwhile to note that the maximum value $\OneThird$ of the visibility of polarization correlations is lower than the bound of $\tfrac{1}{\sqrt{2}}$ in local realistic theories required to satisfy standard Bell's inequalities. This illustrates that the semiclassical description of optical fields is only a particular case in the general class of local realistic theories.

From a more general perspective, the gap between classical and quantum theories calls for a deeper understanding of the role of quantumness in information protocols based on quantum interference and entanglement. This research direction has been shaped by Krzysztof W\'{o}dkiewicz in his last works \cite{CaveWodkPRL04,DemkKusPRA04,WodkHerlPRA98,PraxWasyPRL07}. One intriguing problem is to what degree and under what assumptions classical theories can describe quantum protocols for information processing, especially when one uses input states that have classical counterparts. This issue has been analyzed thoroughly for the case of continuous-variable quantum teleportation \cite{CaveWodkPRL04}, which motivates looking into other scenarios, one interesting candidate being quantum cloning of spin-coherent states \cite{DemkKusPRA04}. Furthermore, despite dramatic differences between predictions of quantum mechanics and classical theories, it is possible to draw insightful parallels between classical and quantum interference phenomena \cite{WodkHerlPRA98}, leading to feasible experimental observations in the optical domain \cite{PraxWasyPRL07}. This opens up a question whether one can find ways to employ classical interference in some applications that are currently thought to depend indispensably on quantum resources.

\section*{Ackowledgments}

We acknowledge the financial support of the Future and Emerging Technologies (FET) programme within the Seventh Framework Programme for Research of the European Commission, under the FET-Open grant agreement CORNER no.\  FP7-ICT-213681. We thank Michael Raymer for his comments.

\end{document}